\documentclass[referee]{jaa} 
\usepackage[english]{babel}
\usepackage{graphics,graphicx,color,amsmath,epsfig,caption,subcaption,float,amssymb}

\begin{document}

\title[FRB event rate]{FRB event rate predictions for the Ooty Wide Field Array} 
\author[ Bhattacharyya et al.] {Siddhartha
Bhattacharyya$^{1}$ \thanks{Email:siddhartha@phy.iitkgp.ernet.in},
Apurba Bera$^{3}$\thanks {Email:apurba@ncra.tifr.res.in}, \newauthor
Somnath Bharadwaj$^{1,2}$,
N. D. Ramesh Bhat $^{4,5}$
\newauthor and Jayaram N. Chengalur$^{3}$
\\ 
$^{1}$ Department of Physics, Indian Institute of Technology,Kharagpur, 721302, India \\
$^{2}$ Centre for Theoretical Studies, Indian Institute of Technology,
Kharagpur, 721302 , India \\ $^{3}$ National Centre for Radio
Astrophysics, Tata Institute of Fundamental Research, Pune 411007,
India\\ $^{4}$ International Centre for Radio Astronomy Research,
Curtin University, Bentley, WA 6102, Australia\\ $^{5}$ Australian
Research Council Centre of Excellence for All-Sky Astrophysics
(CAASTRO)} \date {} \maketitle

\begin{abstract}
We developed a generic formalism to estimate the event rate 
and the redshift distribution of Fast Radio Bursts (FRBs) in our 
previous publication (Bera et al. 2016), considering FRBs are of 
an extragalactic origin. In this paper we present (a) the predicted 
pulse widths of FRBs by considering two different scattering models, 
(b) the minimum total energy required to detect events, 
(c) the redshift distribution and (d) the detection rates of FRBs for 
the Ooty Wide Field Array (OWFA). The energy spectrum of FRBs is 
modelled as a power law with an exponent $-\alpha$ and our analysis 
spans a range $-3\leq \alpha \leq 5$. We find that OWFA will be 
capable of detecting FRBs with $\alpha\geq 0$. The redshift distribution 
and the event rates of FRBs are estimated by assuming two different energy 
distribution functions; a Delta function and a Schechter luminosity 
function with an exponent $-2\le \gamma \le 2$. 
We consider an empirical scattering model based on pulsar observations
(model I) as well as a theoretical model (model II) expected for the
intergalactic medium. The redshift distributions 
peak at a particular redshift $z_p$ for a fixed value of $\alpha$, 
which lie in the range $0.3\leq z_p \leq 1$ for the scattering model I 
and remain flat and extend up to high redshifts ($z\lesssim 5$) for the
scattering model II.
\end{abstract}


\section{Introduction}

Fast Radio Bursts (FRBs) are highly-dispersed, millisecond
duration pulses, discovered at the Parkes radio telescope 
(Thornton et al. 2013 \& Lorimer et al. 2007). The high dispersion measures 
(DM) of the detected FRBs often show $\sim 5-20$ times excess DMs
compared to what is expected from the model for the electron 
density distribution in our Galaxy (Cordes \& Lazio 2002), thereby 
strongly suggesting they are extragalactic phenomena. The observed 
dispersion and scattering indices are believed to be due to propagation 
through cold ionized plasma of the interstellar medium (ISM) 
of our Galaxy, the host galaxy of the FRB and the 
intergalactic medium (IGM). If they indeed originate from cosmological 
distances, FRBs arise in extremely energetic sources with energies 
$\sim10^{33}-10^{35}\;{\rm J}$ being released over
timescales of a few milliseconds.

A total of $26$ FRBs have been discovered over the past 
decade, of which $17$ are reported in the published literature 
(Petroff et al. 2016). Of these $15$ have been detected at 
the Parkes radio telescope and one each at the Arecibo 
radio telescope and the Green Bank radio telescope (GBT). 
One of the FRBs has been found to repeat 
(Spitler et al. 2016), where $10$ detections from the same source have 
been recorded by two different observations separated by about two weeks. This 
suggests that there could be more than one type of progenitors for FRBs.

There are several proposed models for the emission mechanism of FRBs 
(Kulkarni et al. 2015), but as of yet there is no consensus as to which 
is the most likely scenario. The energy spectrum and the energy distribution of 
the events are also not well established. Estimates of the spectral index 
are available only for a few FRBs, but even then they are not reliably
estimated given the poor localization possible with single-dish
observations. A high positive spectral index in the range $7$ to $11$ is 
observed for FRB $121102$ (Spitler et al. 2014) and a high 
negative spectral index ($-7.8$) is observed for FRB $110523$ 
(Masui et al. 2015). Furthermore, a wide range of variation of 
spectral index ($-10.4$ to $13.6$) has been observed for the repeated 
burst (Spitler et al. 2016). Moreover, two FRBs have been detected 
with circular polarization (Petroff et al. 2016).

In our previous work Bera et al (2016), we developed a 
generic formalism to estimate the event rates and the redshift
distribution of FRBs detected by a given radio telescope.
We considered two different scattering models to estimate the observed pulse 
width ($w$) and a power law $E_{\nu}\propto\nu^{-\alpha}$ for 
the energy spectrum of FRBs, where $\nu$ is the observing frequency. 
The energy distribution of FRBs is still unknown and we considered 
two possible energy distribution functions. 
For the purpose of this work, we consider FRBs to be 
one-off transient events (ignoring the repeating FRB, which is clearly
an exception), which can be characterized by a single energy distribution.
The model is normalized by considering FRB $110220$ as the reference event, 
which is the second brightest event (after FRB $010724$, the so-called 
Lorimer burst) detected at the Parkes radio telescope and well characterized 
in terms of its dispersion measure, scattering and fluence properties. 
The estimated total energy ($E_0=5.4\times 10^{33}\;{\rm J}$) of this FRB 
spectrum is considered as the reference energy. We however note that
the value of $E_0$ estimated by using our model (Bera et al. 2016) 
with $\alpha=1.4$, which is differed from the energy estimate 
given by Thornton et al. (2013) by a factor of $5$.

As described in Bera et al. (2016) we consider the detection of four FRBs 
of Thornton et al. (2013) to estimate the event rates. In our formalism 
all redshitfs are inferred from the observed DMs of the reported FRBs by 
considering a host galaxy contribution that is similar to the Milky Way. We 
found that the scattering time scale places an upper limit to redshift 
($z_c$), up to which FRBs can be detected. We estimated the redshift 
distribution of FRBs for the Parkes radio telescope as well as
FRB event rates for the three systems of the Ooty Wide Field Array with 
an incoherent beam formation.

This paper is a follow-up of our previous work (Bera et al. 2016)
and we present a detailed analysis that is applicable 
for the upcoming Ooty Wide Field Array. The Ooty Wide Field Array (OWFA) 
will be a majorly upgraded version of the Ooty Radio Telescope, a 
$530{\rm m}\times30{\rm m}$ parabolic cylinder with a linear feed array of 
dipole receptor elements at its focus, operating at a central frequency of 
$326.5\;{\rm MHz}$. The signals from the individual elements can be combined 
and beam formed in two different ways, incoherent beam formation and a
coherent single beam formation. In the case of coherent single beam 
formation (CA-SB), the voltage signals from the individual elements 
with phase are added directly and then squared to obtain total power. 
The field of view (FoV) is proportional to $\lambda/D$, where $\lambda$ 
is the wavelength of the observation and $D$ is the length of the largest 
baseline. Here the sensitivity is increased by a factor of $N_A$ 
compared to the sensitivity achieved by a single element. 
$N_A$ is total number of elements, where an element is a segment spanning 
$24$ dipole antennas for OWFA Phase I and $4$ dipole antennas for OWFA Phase
II. In the case of incoherent beam formation (IA), the squares of the 
voltages from the individual elements are summed over to 
obtain the total power. This mode of beam formation does not contain any 
phase information. Here the FoV is proportional to $\lambda/d$, where 
$d$ is the length of a single element. Note that, for 
parabolic dishes the field of view is proportional to $\lambda/D_P$,
where $D_P$ is the diameter of the parabolic dish, whereas for cylindrical 
reflectors, the field of view is $\pi\times\lambda/b\times\lambda/d$, 
where $b$ and $d$ are the dimensions along the parallel and perpendicular 
to the cylindrical axis respectively.
In this case the sensitivity is increased by a factor of $\sqrt{N_A}$ 
compared to the sensitivity achieved by a single element. 
More information about the baselines of OWFA Phase I and Phase II can be 
found in Ali \& Bharadwaj (2014).

We also consider here a third kind of beam formation, which we call 
coherent multiple beam formation, which is the mixture of IA and CA-SB 
mentioned above. In coherent multiple beam formation (CA-MB), one forms 
the IA to obtain a large instantaneous field of view but at a relatively
shallow sensitivity. When an event is detected in 
the IA mode,the high time resolution signals are recorded to eventually 
form multiple coherent beams offline in all possible directions.
This will give us the sensitivity of the CA-SB, but with the field of 
view of the IA. This specific kind of strategy was first demonstrated in a 
pilot transient survey with the Giant Metrewave Radio Telescope (GMRT) by 
Bhat et al. (2013).

In this paper, we estimate the FRBs event rates followed by the predicted 
pulse widths, the minimum total energy required to detect events and the 
redshift distribution of FRBs for the OWFA. We then compare our results 
with another two cylindrical radio telescopes that are gearing up for FRB 
events, UTMOST and CHIME. UTMOST (Caleb et al. 2016) is a cylindrical radio 
telescope that operates at an observational frequency of $843\;{\rm MHz}$ 
whereas CHIME (Newburgh et al. 2014) is an upcoming radio interferometer,
which will operate at an observational frequency $600\;{\rm MHz}$.

We provide estimates for the detection rates for all three beam forming 
modes for OWFA along with UTMOST and CHIME. However, we are 
considering only the IA and CA-SB modes for detailed comparisons, 
as a realistic implementation of CA-MB modes (for OWFA) will be largely 
dictated by affordable computational resources. As mentioned earlier, 
in the case of OWFA, we 
also consider the possibility, where we get to take advantage of 
both wide FoV and the full sensitivity, i.e. the use of IA for wide 
FoV (at the expense of shallow searches), but then will go for 
high-quality signal detection via CA-MB for promising candidates. 
We however clarify that this is not equivalent to an all-time CA-MB 
search.

A brief outline of the paper is as follows. In Section 2, we present 
a short description of the Ooty Wide Field Array along with UTMOST and CHIME. 
The predictions for the FRB event rates for OWFA, UTMOST and CHIME 
are described in Section 3 and we discuss and summarize our results in 
Section 4.

\section{The Ooty Wide Field Array} 
The Ooty Wide Field Array (OWFA) is an upgraded 
version of the Ooty Radio Telescope (ORT) that was built in early 70's 
(Swarup et al. 1971). ORT has a long cylindrical reflector of dimension 
$530{\rm m}\times 30{\rm m}$. It contains $1056$ half wavelength linear 
dipoles along the focal line of the parabolic cylindrical reflector. 
Therefore, it is sensitive only to a single polarization component of 
the incident radiation. The signals from these dipoles are currently 
combined, using an analogue beam forming network and we refer to this as 
the Legacy System. The signals from all $1056$ dipoles are 
essentially combined to effectively form a coherent beam.

The Legacy System operates at an observational frequency 
$\nu_0=326.5\,{\rm MHz}$ ($\lambda=0.91\,{\rm m}$) with bandwidth 
$B=4\,{\rm MHz}$. The ongoing upgrade envisages two modes of operation; 
Phase I \& Phase II. In Phase I, each set of $24$ dipoles will be 
combined to form a single element, whereas in Phase II, each element 
will consist of $4$ dipoles. Only $40$ of the $44$ such
sets (half-modules) will be used in Phase I, whereas Phase II will 
make use of all $1056$ dipoles in the form of $264$ elements.
The bandwidth of Phase I and Phase II are $19.2\,{\rm MHz}$ 
and $38.4\,{\rm MHz}$ respectively, centred at the same observational 
frequency of the ORT Legacy System. More technical information about OWFA 
can be found in Subrahmanyan et al. (2016). The parameters of OWFA that are
relevant for this work are tabulated in Table 1.

In this work the primary beam patterns $B(\vec{\theta})$ of the 
current and the new phases of OWFA (Ali \& Bharadwaj 2014) are assumed 
to be given by

\begin{equation} 
B(\vec{\theta})={\rm sinc}^2\left(\frac{\pi d\theta_x}{\lambda}\right) 
{\rm sinc}^2\left(\frac{\pi b\theta_y}{\lambda}\right)
\label{beam_pattern}
\end{equation}

where, $b \times d$ is the aperture of a single element and $(\theta_x,\theta_y)$ 
are components of $\vec{\theta}$ on the plane of the sky. Two other 
radio telescopes, UTMOST (Caleb et al. 2016) and CHIME (Newburgh et al. 2014), 
are quite similar to OWFA, comprising long cylindrical reflectors and 
operating at a single radio frequency ($843\,{\rm MHz}$ and $600\,{\rm MHz}$, 
respectively), though CHIME will be equipped to record data over a large bandwidth. 
The parameters of UTMOST and CHIME are also tabulated in Table 1.

\begin{table}
\begin{tabular}{|c|c|c|c|c|c|c|c|c|} 
\hline
Telescope & Beam & $N_A$ &$\nu$& $B$ & $\Delta\nu_{c}$ & $A_S$ & FoV & ${\rm FoV} \times A_S$ \\
 & Formation & & (MHz) & (MHz) & (kHz) & (Jy$^{-1}$) & (deg$^2$) & (deg$^2$ Jy$^{-1}$) \\
\hline 

ORT Legacy & $-$ & $1$ & & $4$ & $125$ & $0.023$ & $0.18$ & $0.004$ \\
System & & & & & & & & \\
\cline{1-3} \cline{5-9}

 & IA & & & & & $0.004$ & $8.05$ & $0.032$\\
\cline{2-2} \cline{7-9}
OWFA & CA-SB & $40$ &  & $19.2$ & $24$ & $0.02$ & $0.21$ & $0.004$\\
\cline{2-2} \cline{7-9}
Phase I & CA-MB & & $326.5$ & & & $0.02$ & $8.05$ & $0.161$\\
\cline{1-3} \cline{5-9}

 & IA  &  & & & & $0.001$ & $47.93$ & $0.048$\\
\cline{2-2} \cline{7-9}
OWFA & CA-SB & $264$ &  & $38.4$ & $48$ & $0.022$ & $0.18$ & $0.004$\\
\cline{2-2} \cline{7-9}
 Phase II & CA-MB & & & & & $0.022$ & $47.93$ & $1.054$\\
\hline

 & IA & & & & & $0.003$ & $7.8$ & $0.023$\\
\cline{2-2} \cline{7-9}
UTMOST & CA-SB & $352$ & $843$ & $31.25$ & $781.25$ & $0.057$ & $0.07$ & $0.004$\\
\cline{2-2} \cline{7-9}
 & CA-MB & & & & & $0.057$ & $7.8$ & $0.445$\\
\hline

 & IA &  & & & & $0.003$ & $132$ & $0.4$\\
\cline{2-2} \cline{7-9}
CHIME  & CA-SB & $1280$ & $600$ & $400$ & $1000$ & $0.099$ & $0.29$ & $0.029$\\
\cline{2-2} \cline{7-9}
 & CA-MB &  & & & & $0.099$ & $132$ & $13.1$\\
\hline
\end{tabular}

\caption{We estimate the sensitivity of the telescope $A_S$ by 
considering system temperatures of OWFA, UTMOST and
CHIME as $150\,{\rm K}$, $70\,{\rm K}$ and $50\,{\rm K}$ respectively.
We consider here three kinds of beam formation, incoherent beam formation
(IA), coherent single beam formation (CA-SB) and coherent multiple beam 
formation (CA-MB). The symbols $N_A$, $\nu$, $B$, $\Delta\nu_{c}$ and FoV 
stand for number of elements, observational frequency, bandwidth, spectral 
channel width and field of view of the telescope. The efficiency of a 
single element $\eta=0.6$ is assumed to be same for OWFA, UTMOST and CHIME.}
\end{table}

The detection probability of FRBs largely depends on two factors, 
the field of view (FoV) and the sensitivity ($A_S$) of the 
telescope\footnote{The sensitivity $A_S$ is defined as the ratio 
of the gain $G$ of a single element to the system temperature 
$T_{sys}$ of the telescope}. A telescope with a large field of view 
and a high sensitivity will be capable of detecting FRBs in large numbers. 
Therefore, the product ${\rm FoV} \times A_S$ is a useful indicator 
of the detection prospects of FRBs for any given radio telescope.
This product is maximum for CHIME in the CA-MB mode, however the three 
telescopes operate at different frequencies and therefore highly complementary. 
All these three telescopes also have effectively higher FRB detection 
sensitivity than the Parkes radio telescope or the Arecibo 
radio telescopes. For example, for the Parkes radio telescope, 
${\rm FoV} \times A_S=8.94\times 10^{-4}\;{\rm deg}^2{\rm Jy}^{-1}$, 
which is $\sim 10^3$ times smaller than that of OWFA in the CA-MB mode.

\section{Predictions of FRB event rates}
In this section, we briefly describe the method we use to predict the 
detection rates of FRBs followed by the estimation of the predicted 
pulse widths, minimum total energy required to detect events and the 
redshift distribution of FRBs for a given radio telescope. 

Further details on the formalism employed are reported in our previous 
work Bera et al. (2016).

\subsection{Pulse width}
The observed pulse width $w$ of a FRB at a redshift $z$ with an intrinsic 
pulse width $w_i$ is given by 

\begin{equation}\label{w_obs} 
w = \sqrt{w_{cos}^2+w_{DM}^2+w_{sc}^2}
\end{equation}

where, $w_{cos}$, $w_{DM}$ and $w_{sc}$ are the contribution from the 
cosmic expansion, dispersion broadening and scatter broadening, respectively. 
The term $w_{cos}$ arises here due to the cosmological expansion of the 
universe, which is the product of $w_i$ and $(1+z)$. The cold ionized 
medium of the ISM in our Galaxy as well as the host galaxy of the FRB 
and the IGM introduces dispersion smearing and scatter broadening. The 
observed radio signal is assumed to be incoherently de-dispersed. 
This would leave a residual dispersive smearing corresponding to the 
channel width, viz. $w_{DM}$. The exact mechanism of the scattering in 
the intervening medium is still unknown and we consider here two different 
scattering models; the ones based on, Bhat et al. (2004) and Macquart 
\& Koay (2013), and denote them as the scattering model I (Sc-I) and 
the scattering model II (Sc-II), respectively. 

The scattering model I is an empirical fit to a large body of pulsar 
measurements in the Milky Way, which we have rescale it for
the intergalactic medium, and given by

\begin{equation}\label{sc1}
\log\:w_{sc}=C_0 + 0.15\:\log\:DM_{IGM}+1.1\:(\log\:DM_{IGM})^2-3.9\log\:\nu
\end{equation} 

where $C_0$ is a normalization constant and we have used $C_0=3.2$ to 
rescale the scattering model for the intergalactic medium. $DM_{IGM}$ and
$\nu$ are the dispersion measure due to the intergalactic medium and the 
observational frequency of the radio telescope respectively. 
 
The scattering model II is largely a theoretical framework by
considering the turbulence in the intergalactic medium, and given by

\begin{equation}\label{sc2}
w_{sc}(z)=  \frac{k_{sc}}{\nu^4 Z_L} \int_0^z  D_H(z')dz' \int_0^z (1+z')^3 D_H (z') dz'
\end{equation}

where, 
\begin{align*}
&D_H(z)= (\Omega_{m}(1+z)^3+\Omega_{\Lambda})^{-1/2}, \\
&Z_L= (1+z)^2 \left[(1+z)-\sqrt{z(1+z)}\right]^{-1}
\end{align*}
and we have used the normalization constant 
$k_{sc}=8.5\;\times\;10^{13}\;{\rm ms\;{MHz}^4}$.
Both normalization constants $C_0$ and $k_{sc}$ are estimated by considering
the pulse width of the reference event FRB $110220$, which is $w=5.6\;{\rm ms}$
at redshift $z=0.8$ with an intrinsic pulse width $w_i=1\;{\rm ms}$.
Note that in Eqs. \ref{sc1} and \ref{sc2}, we have used $\nu_0$ in 
${\rm MHz}$, $w_{sc}$ in ${\rm ms}$ and $DM$  in ${\rm pc}\,{\rm cm}^{-3}$.

\begin{figure}[h]
\label{fig-1:pulsewidth}
\centering \input{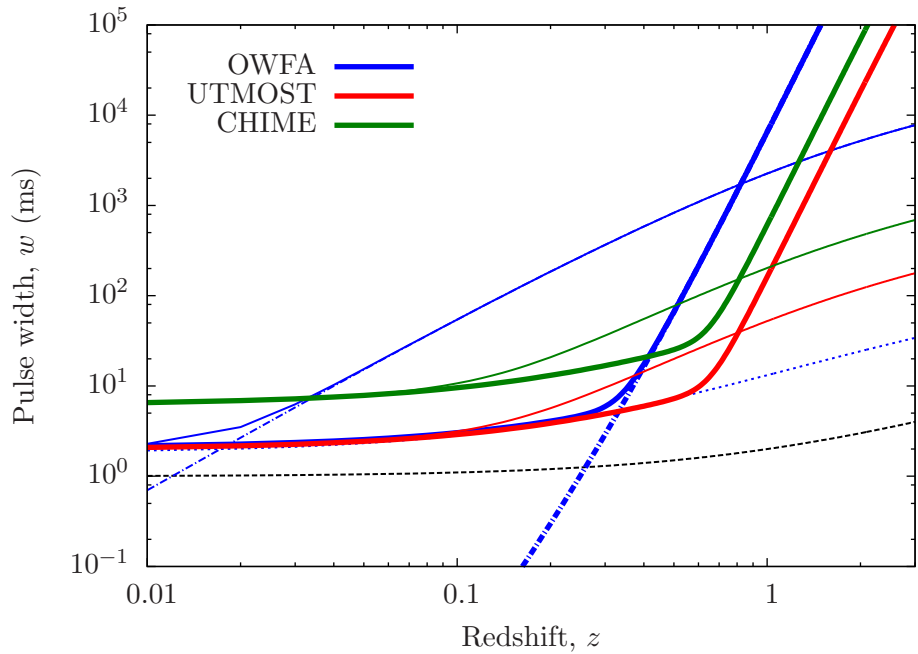}
\caption{The predicted pulse width of a FRB located at a redshift $z$ 
with intrinsic pulse width $w_i=1\;{\rm ms}$ (see text for details), assumed 
to be detected by using OWFA (Blue), UTMOST (Red) and CHIME (Green). The black 
dashed line refers the cosmic expansion, which is independent of the telescope
configuration. The dotted and dot-dashed lines refer the dispersion broadening 
and the scatter broadening respectively. The thick and thin lines correspond to 
scattering model I (Sc-I) and scattering model II (Sc-II) respectively.}
\end{figure}

Figure 1 shows $w$ as a function of $z$ for FRB detection using OWFA. 
The different components of $w$ for the two scattering models and the predicted 
values of $w$ for UTMOST and CHIME, are also shown in this figure. 

Contribution from dispersion smearing can be significant at higher DMs; for 
example, for $DM=1000\;{\rm pc\;cm}^{-3}$, we find that the values 
of $w_{DM}$ are $29.78\;{\rm ms}$, $5.72\;{\rm ms}$ and $11.44\;{\rm ms}$ 
for Legacy System, OWFA Phase I and Phase II, respectively. These 
however contribute $\lesssim 1\%$ to the predicted pulse width of FRBs for OWFA. 
Therefore, $w_{DM}$ makes negligible contribution to $w$ but $w_{sc}$ 
dominates at large $z$. As discussed above, $w_{DM}$ depends on the 
channel width of the detection backend (spectrometer), whereas $w_{sc}$ 
solely depends on the observational frequency. Therefore, Fig. 1 is 
similar for all different modes of OWFA. The pulse width is independent 
of the rms noise and the field of view of the telescope. Thus, Fig. 1 
is also similar for different modes of beam formation.

Figure 1 shows that the term $w_{cos}$ makes an insignificant
contribution to $w$ for all redshifts for the assumed $w_i=1\;{\rm ms}$. 
For the scattering model I, $w_{sc}$ makes less contribution to $w$ 
at low redshift ($z<0.3$) in comparison with $w_{DM}$ and it starts 
to dominate at $z\geq0.3$. For the scattering model II, $w_{sc}$ is 
very high and dominates $w$ even at very low redshifts ($z\approx 0.03$). 
The value of $w$ for the scattering model II is greater that of the
scattering model I up to $z\approx 0.8$, after which the latter one 
increases very sharply.

The predicted pulse widths for UTMOST and CHIME are smaller than 
these of OWFA, which follows from the fact that they operate at 
higher observing frequencies. However, the qualitative nature of 
the curves are similar to those of OWFA. We have repeated the 
entire analysis with different values of $w_i$ ($0.5\;{\rm ms}$ 
\& $2\;{\rm ms}$). The normalization constants of both scattering 
models differ but the overall changes in $w$ is negligible and 
hence do not affect the curves. It essentially shows that the 
qualitative nature of $w$ is independent of the choice of 
$w_i$ for all redshifts.

\subsection{Minimum Energy}
The detectability of a given FRB depends on the amount of 
total energy received by the telescope from the source. 
The minimum total energy $E_{min}$ required to detect FRBs, is given by

\begin{equation}\label{E_min} 
E_{min}=(9.55\times10^{27})\;\frac{4\pi r^2 F_l}{\overline{\phi}(z)B(\vec{\theta})}\,\sqrt{\frac{w}{1\;{\rm ms}}}
\end{equation}

where $F_l$ is the limiting fluence, which further depends on the 
minimum signal to noise ratio $S/N$ of the observation and the rms 
noise of the telescope for $1\;{\rm ms}$ observation. We consider 
$S/N=10$ here. Here, fluence is defined as the time 
integral of flux density. As FRBs are one-off bursts of radiation,
it is convenient to use fluence to quantify the integrated energy 
over the duration of the burst. 
The predicted pulse width $w$ is scaled to $1\;{\rm ms}$. 
The comoving distance $r$ is estimated considering the standard
$\Lambda{\rm CDM}$ cosmology (Spergel et al. 2003). $B(\vec{\theta})$ 
is the beam pattern, which is calculated using Eq. 
\ref{beam_pattern}. $\overline{\phi}(z)$ is the emission line profile 
averaged over frequency, which is normalized by considering FRB $110220$ 
as the reference event and the estimated total energy 
of the FRB as the reference energy $E_0=5.4\times 10^{33}\;{\rm J}$. 
We have used $F_l$ in ${\rm Jy\;ms}$, $r$ in ${\rm Gpc}$, 
$\nu$ in ${\rm MHz}$ and $E_{min}$ in ${\rm J}$.

\begin{figure}[]
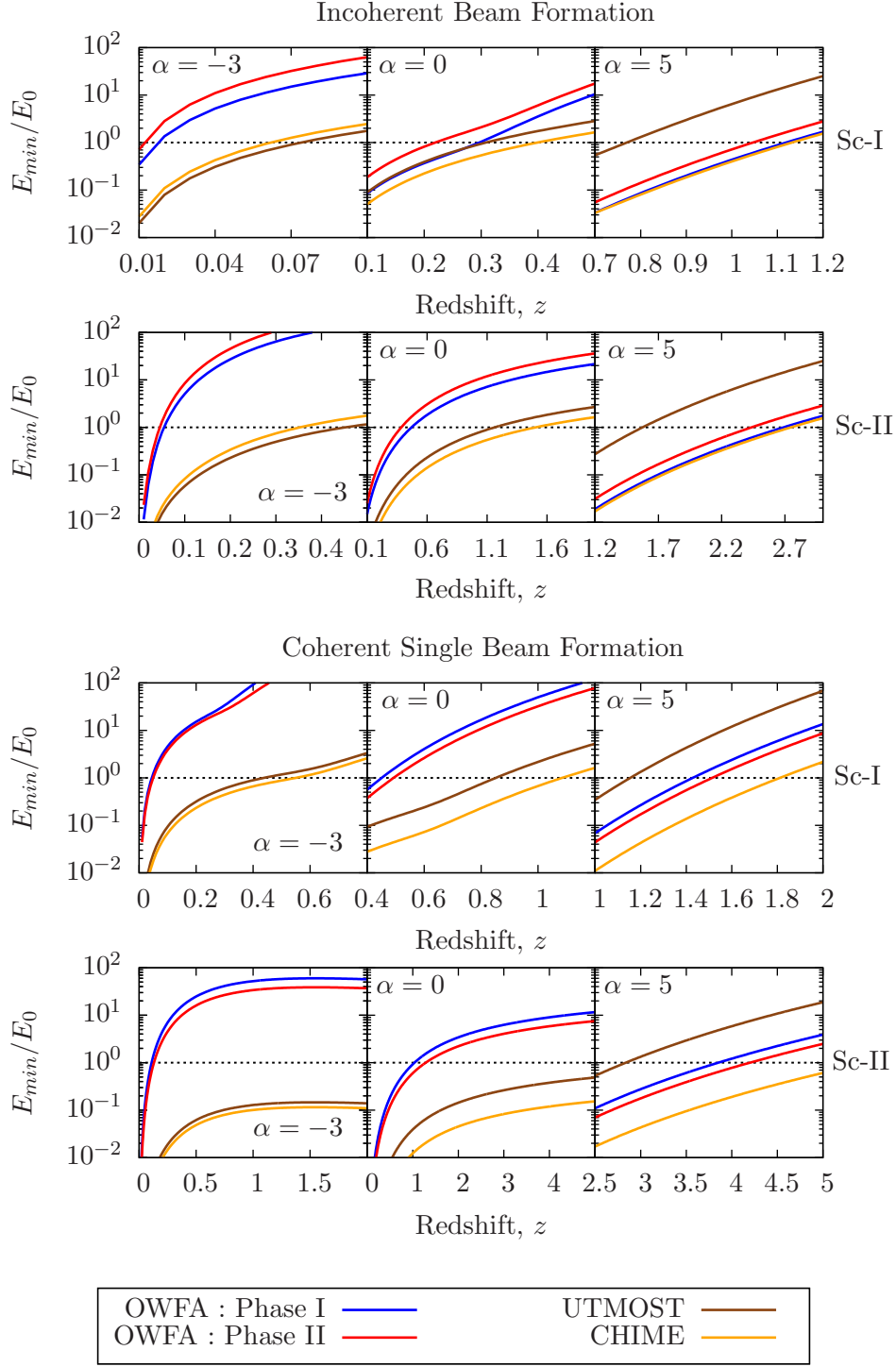

\label{fig-2:energy_ratio}
\begin{center}
\input{energy_ratio_IA.tex}
\input{energy_ratio_CA.tex}
\input{notation_energy_ratio.tex}
\end{center}
\caption{The minimum total energy $E{min}$ required to detect a 
FRB at a redshift $z$ at the beam center of OWFA with its different 
phases and different values of the spectral 
indices $\alpha$, i.e. $S(\nu)\propto\nu^{-\alpha}$, ($-3\leq \alpha \leq 5$) 
in units of reference energy $E_0$. The top and bottom panels 
correspond to incoherent beam formation and coherent Single
beam formation modes. The first and second halves of each panel 
correspond to scattering model I (Sc-I) and scattering model II
(Sc-II) respectively. We consider $w_i=1\;{\rm ms}$.}
\end{figure}

\begin{figure}[]
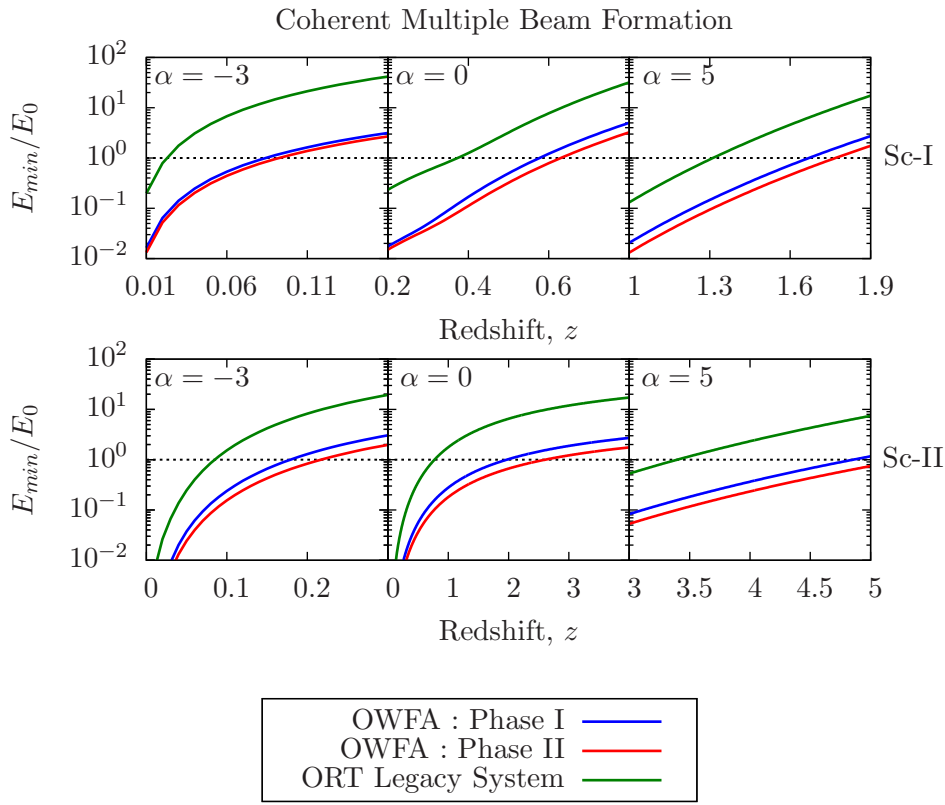

\label{fig-3:energy_ratio}
\begin{center}
\input{energy_ratio_HB.tex}
\input{notation_energy_ratio_1.tex}
\end{center}
\caption{Similar to Fig. 2, but only for OWFA Phase I and 
Phase II with coherent multiple beam formation mode. The green 
line represents $E{min}$ for ORT Legacy System.}
\end{figure}

Figure 2 shows $E_{min}$ as a function of $z$ assuming the detected FRBs 
located at the beam center of OWFA, UTMOST and CHIME, expressed in units of 
$E_0$. The different plots in a row correspond to different values of 
the spectral indices $\alpha$, spanning a large range ($-3\le\alpha\le 5$), 
where the flux $S(\nu)\propto \nu^{-\alpha}$ and a fixed $w_i=1\;{\rm ms}$. 
$E_{min}$ depends on the rms noise of observation 
and the predicted pulse width ($w$) of FRBs (Eq. \ref{E_min}). 
The values of $w$ are the same for all phases of the OWFA with 
different beam formations but the rms noises are different. 
Therefore, the values of $E_{min}$ are also different. Similar variations
are also observed for UTMOST and CHIME. Incoherent and coherent single 
beam formation along with the two scattering models are shown in the top
and bottom panels of the figure. For OWFA, UTMOST and CHIME, the predicted 
pulse widths increase very sharply for the scattering model I in comparison 
to the scattering model II, hence $E_{min}$ also increases accordingly.

Figure 3 shows the results for OWFA Phase I and Phase II with coherent 
multiple beam formation and ORT Legacy System. The qualitative nature of 
Fig. 3 and Fig. 2 are similar. The dotted horizontal lines in both 
the figures are for $E_{min}=E_0$. Assuming that all FRBs have equal 
total energy $E\approx E_0$, the intersection of the curves 
with the dotted lines gives a useful indication of the redshift up to which 
the telescope is sensitive enough to detect FRBs. This is denoted here as 
the cut off redshift $z_c$.

We first discuss the results for the incoherent beam formation mode, as shown 
in Fig. 2. For $\alpha=-3$, the cutoff redshifts are typically small for 
the scattering model I, while for scattering model II, the cutoff redshifts 
are quite feasible for both UTMOST and CHIME. For $\alpha=5$, the curves 
corresponding to CHIME and OWFA Phase I overlap for both the scattering models. 
We find here that the values of $z_c$ for UTMOST are small in comparison to 
others for both the scattering models. Coherent single beam formation, with 
$\alpha=5$, also exhibits a similiar feature.

Now we discuss the results for coherent single beam formation mode as shown 
in Fig. 2. For the scattering model I, the nature of the curves are quite 
similar to that of incoherent beam formation. For the scattering model II 
with $\alpha\le0$, we do not have any cut off redshift for UTMOST and CHIME. 
It shows that both these telescopes can in principle probe FRBs out to 
arbitrarily high redshifts. Figure 3 shows that the cut off redshifts for 
the ORT Legacy System are always small compared to others. The other features 
of Fig. 3 are quite similar to Fig. 2. For the rest of the analysis we 
ignore the negative values of $\alpha$, as it is consistent with currently 
available observational constraints. We repeated the entire analysis with 
$w_i=0.5\;{\rm ms}$ and $2\;{\rm ms}$ and find that at low redshift ($z\ll z_c$), 
there is a small deviation between the curves but overall there is no change 
in $z_c$ (though not shown in Fig. 2 and Fig. 3). Therefore, henceforth, 
we shall use the value of intrinsic pulse width $w_i=1\;{\rm ms}$.

\subsection{Redshift Distribution}

We now consider a quantity, $N_{det}$, which is total number of events 
expected to be detected for an observation time $T$. This quantity is 
obtained by integrating the comoving number density of FRBs $n(E,w_i,z)$ 
over energy, pulse width, the solid angle subtended by the telescope 
beam and redshift. The quantity $N_{det}$ is given by

\begin{equation}\label{N_det}
 N_{det}(T)= T \int dz \frac{dr}{dz}
\left(\frac{r^2}{1+z}\right) \int d\Omega\int dw_i
\int_{E_{min}(z)}^{\infty} dE\:\:n(E,w_i,z)
\end{equation}

where, $n(E,w_i,z)$ is the comoving number density of FRBs. 
The redshift integral is going up to the cut off redshift $z_{cut}$.
$N_{det}$ depends on the beam pattern of the telescope. We assume the 
sampling time of the telescope to be $\le 1\;{\rm ms}$ so that the FRB 
signal is resolved. The cut off redshift $z_c$ does not depend on the 
intrinsic pulse width $w_i$ of the source and we assume here that all 
FRBs have the same intrinsic pulse width $w_i=1\;{\rm ms}$.
We further assume that $n(E,w_i,z)$ does not evolve with redshift, so 
$n$ is solely a function of $E$. As the emission mechanism and the energy 
distribution of FRBs remain unknown, we consider two possible energy 
distribution functions, viz. a Delta function and a Schechter luminosity 
function. For the Delta function, we assume that all FRBs have 
the same energy that is equal to the reference energy 
$E_0=5.4\times10^{33}\;{\rm J}$, whereas in the Schechter luminosity function, 
an energy spread is allowed, is given by,

\begin{equation}
n(E,w_i,z)=\frac{n_0}{E_0}\left(\frac{E}{E_0}\right)^{\gamma}\exp\left(-\frac{E}{E_0}\right)
\end{equation}

where the normalization constant $n_0$ is a free parameter, which is
estimated by considering the detection of four FRBs during an effective
observation time of $298$ days with a single beam of the Parkes 
radio telescope (Thornton et al. 2013).
We consider the exponent $\gamma$ in the range $-2\le \gamma \le 2$. 
For a negative exponent ($\gamma < 0$), we fix the lower cut off of the energy as 
$E_0/100$ to normalize the function. The redshift distribution of FRBs 
refers to the variation of fraction of predicted events 
($\Delta N_{det}/N_{det}$) with redshift $z$. Here $\Delta N_{det}$ is 
the number of events, which are expected to be detected over a redshift bin 
($z_{bin}$) and $N_{det}$ is the total number of events, expected to be 
detected over a redshift range ($0\le z \le z_{max}$). We consider 
$z_{bin}=0.1$ and $z_{max}=5$.

\begin{figure}
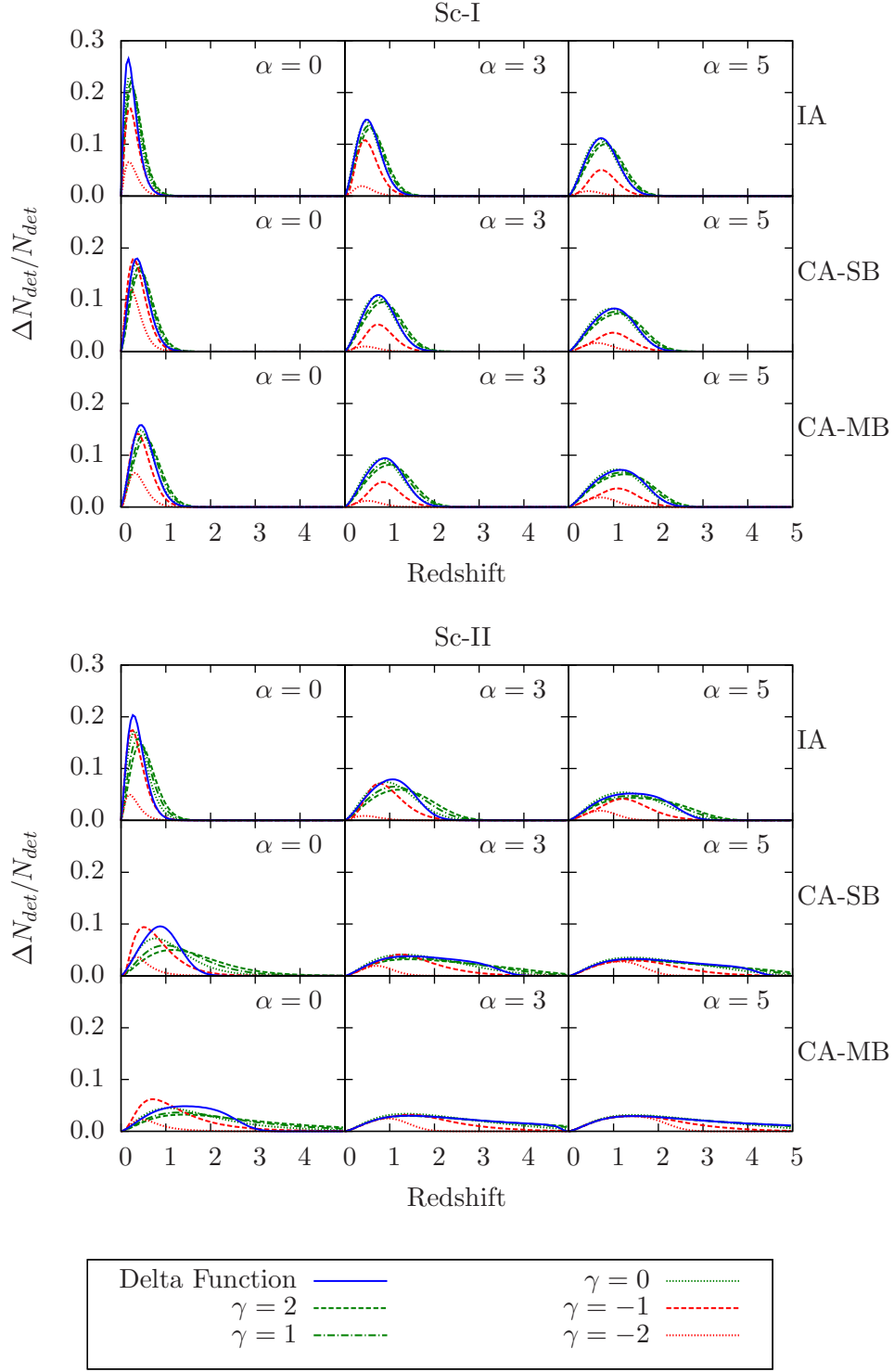

\label{fig-4:redshift_distribution}
\centering 
\input{redshift_dist_owfa_sc1.tex}
\input{redshift_dist_owfa_sc2.tex}
\input{notation_redshift_dist.tex}
\caption{The fraction of predicted FRBs over the bin on redshift axis 
($z_{bin}=0.1$) for OWFA Phase II with different beam formation modes 
and different energy distribution functions of FRBs. Different 
panels in a row correspond to different values of $\alpha$ 
($0\le \alpha \le 5$). Upper and lower halves correspond to 
scattering model I (Sc-I) and scattering model II 
(Sc-II) respectively with $w_i=1\;{\rm ms}$.}

\end{figure}

Figure 4 shows $\Delta N_{det}/N_{det}$ as a function of $z$
with $z_{bin}=0.1$ and $0\le \alpha \le 5$ for OWFA Phase II 
with different beam formation modes and energy distribution 
models of FRBs. 
The upper and lower halves correspond to the scattering models I 
and II, respectively, with $w_i=1\;{\rm ms}$.
The area under the curves gives an indicator of the total number of 
predicted FRBs for a particular beam formation mode of the OWFA 
with a particular energy distribution function of FRBs and a particular
value of $\alpha$. Figure 3 is very similar for all phases
of OWFA, UTMOST and CHIME and are hence not shown here.

We find that the curves peak at a particular value of redshift 
$z_p$. The values of $z_p$ increases with increasing $\alpha$ 
($0\le \alpha \le 5$). The stiffness and the peak values of the 
curves are dissimilar for different beam formation modes and for different
scattering models. In the case of scattering model I, for the values of 
$\alpha=0$, $3$ and $5$, the approximate value of
$z_p$ are $0.3$, $0.7$ and $1$ respectively. The curves corresponding 
to the Delta function and the Schechter luminosity function with a
positive exponent ($\gamma\ge0$), are almost overlapping, whereas the 
curves for a negative exponent ($\gamma<0$) are different. The area under 
the curves of the Schechter luminosity function with $\gamma=-2$ are small 
and the corresponding peak values are low. The highest peak value of
the curves is maximum ($\sim 0.25$) for the IA beam formation mode with 
$\alpha=0$ and minimum ($\sim 0.08$) for CA-MB beam formation mode 
with $\alpha=5$. 
For the scattering model II, Fig. 3 shows that we can expect to detect 
FRBs at higher redshifts but we estimate a cut off $z=5$. Here all 
the plots are similar, only the plots for IA beam formation mode with 
$\alpha=0$ are slightly different. The curves are almost flat with 
low peak values for $\alpha>0$. The curves corresponding to different 
energy distribution functions of FRBs are overlapping with each other. 
The highest peak value of the curves and the area under the curves 
for the Schechter luminosity function with $\gamma=-2$, are similar 
to those of the scattering model I.

\subsection{Event rate}
Finally, we consider FRB detection rates for a telescope with 
given parameters. We use Eq. \ref{N_det} to estimate this 
quantity for the OWFA, UTMOST and CHIME; in each case for 
different beam formations. We use the detection rates 
from Thornton et al. (2013) as our reference.

\begin{figure}[]
\label{fig-5:event_rate_IA}
\centering 
\input{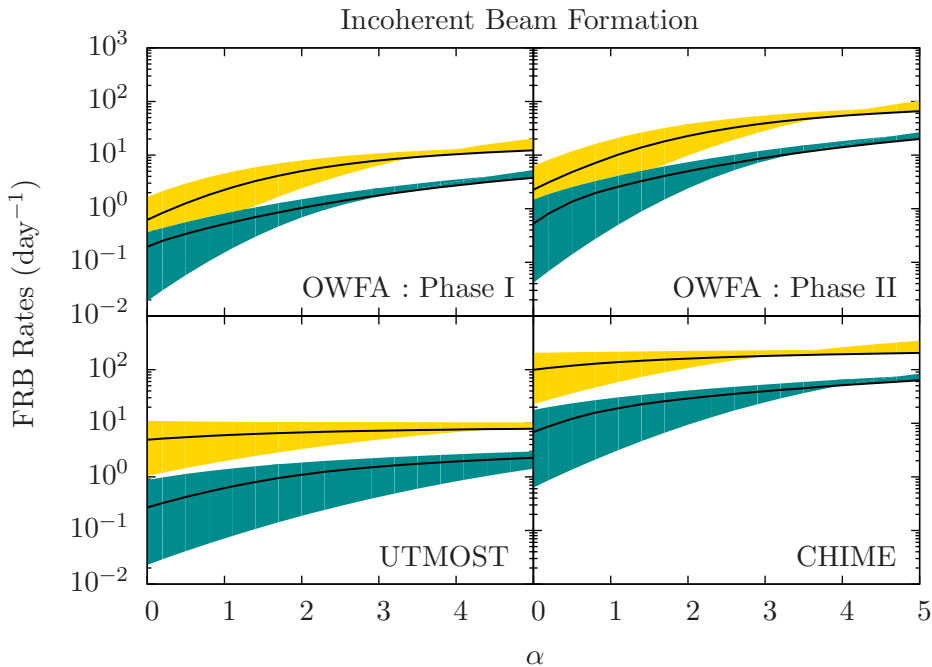}
\caption{The number of FRBs expected to be detected per day by using OWFA, 
UTMOST and CHIME with incoherent beam formation. The yellow and 
green regions correspond to scattering model I (Sc-I) and 
scattering model II (Sc-II) respectively. Solid black lines 
denote Delta function, while the boundaries of the regions enclose the curves
corresponding to the Schechter luminosity function with 
exponent in the range $-2\le \gamma \le 2$.}
\end{figure}

\begin{figure}[]
\label{fig-6:event_rate_CA-SB}
\centering 
\input{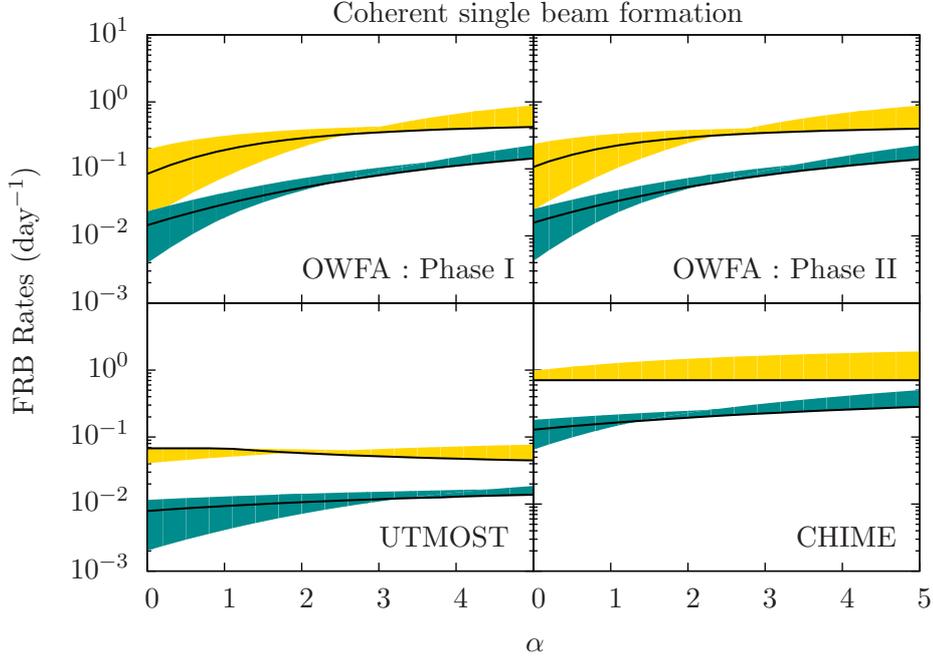}
\caption{Same as Fig. 5 with coherent single beam formation.}
\end{figure}

\begin{figure}[]
\label{fig-6:event_rate_CA-MB}
\centering 
\input{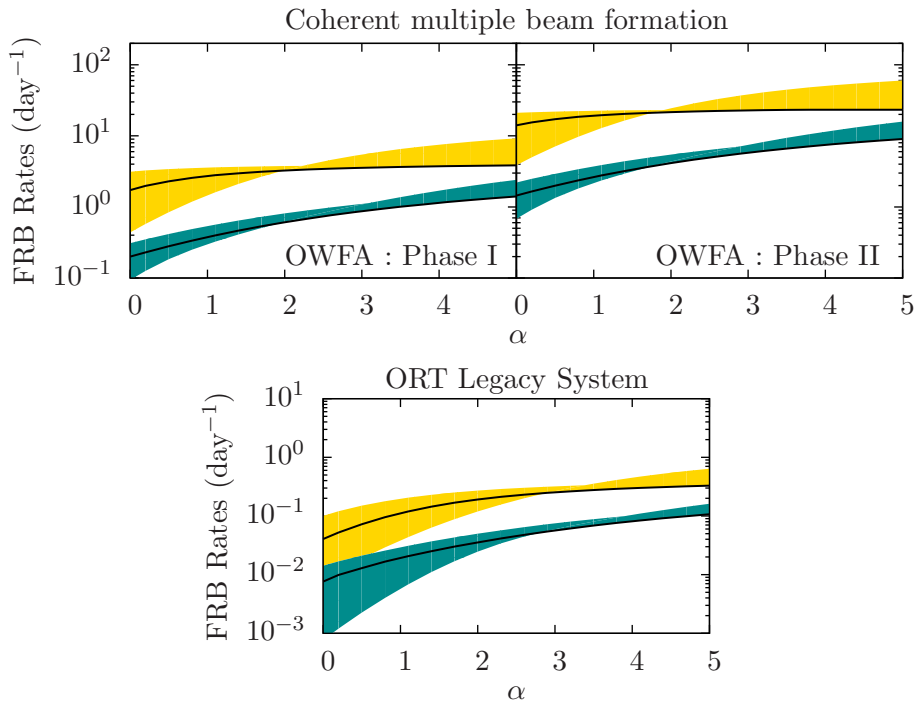}
\caption{Same as Fig. 5. The upper panels denote OWFA Phase I and
Phase II with coherent multiple beam formation. The lower panel
denotes ORT Legacy System.}
\end{figure}

Figures 5 and 6 show the FRB detection rates (${\rm day}^{-1}$) as a function 
of $\alpha$ for OWFA, UTMOST and CHIME with incoherent and coherent Single
beam formations, whereas in Fig. 7 we show these rates for OWFA Phase I 
and Phase II, for coherent multiple beam formation and the ORT Legacy System.  
The yellow and green solid regions represent the scattering models I and 
II, respectively. The solid black lines refer to the Delta function 
whereas the boundaries of the regions enclose the curves representing 
the Schechter luminosity function with exponents in the range 
$-2\le \gamma \le 2$. The detection rates (${\rm day}^{-1}$) increase, 
with increasing $\alpha$ ($\alpha>0$). The detection rates corresponding to 
the scattering model I are large in comparison to the that of scattering 
model II. The approximate numbers of FRBs expected to be detected
per day for OWFA, UTMOST and CHIME with the consideration of Delta function
as the energy distribution function of FRBs and $\alpha=1.4$ are tabulated
in Table 2 for comparison.

\begin{table}[]
\begin{center}
\begin{tabular}{|c|c|c|c|} 
\hline
Telescope & Beam      & \multicolumn{2}{c|}{Number of FRBs per day}\\
\cline{3-4}
          & Formation & Scattering & Scattering \\
          &           & Model I    & Model II \\
\hline 

ORT Legacy & $-$ & $0.14$ & $0.03$ \\
System     &     &        &        \\
\hline

           & IA    & $3.25$ & $0.70$ \\
\cline{2-4}
OWFA       & CA-SB & $0.23$ & $0.04$ \\
\cline{2-4}
Phase I    & CA-MB & $14.85$ & $2.31$ \\
\hline

           & IA    & $13.74$  & $3.34$ \\
\cline{2-4}
OWFA       & CA-SB & $0.25$   & $0.04$ \\
\cline{2-4}
Phase II   & CA-MB & $103.12$ & $15.94$ \\
\hline

           & IA    & $6.38$ & $0.81$ \\
\cline{2-4}
UTMOST     & CA-SB & $0.06$ & $0.01$ \\
\hline

           & IA    & $144.65$ & $22.42$ \\
\cline{2-4}
CHIME     & CA-SB  & $0.71$   & $0.17$ \\
\hline

\hline
\end{tabular}
\end{center}

\caption{The approximate number of FRBs expected to be detected per day is estimated 
by considering Delta function as the energy distribution function of FRBs 
with $\alpha=1.4$.}
\end{table}

It is easily shown that a large number of FRBs are likely to be detected 
by OWFA with coherent multiple beam formation rather than coherent 
single beam formation. In particular we expect to detect $\sim 100$ FRBs per 
day with OWFA Phase II in the case of coherent multiple beam formation. The 
detection per day for OWFA Phase II are large compared to the that for 
UTMOST. Table 2 shows that the detection per day is even larger for CHIME 
with incoherent beam formation. This is not surprising since CHIME has a 
large field of view compared to others. However, as we emphasized earlier, 
these three telescopes operate in different frequency ranges and hence 
complementary. The approximate numbers for FRBs detections are summarized 
in Table 2 for the specific case of $\alpha=1.4$ to allow further 
meaningful comparison.

\section{Summary \& Conclusions} 
We have presented a detailed analysis of FRB detection rates for different 
phases of OWFA using a generic formalism prescribed in our previous work 
(Bera et al. 2016).
We compare the results with two other cylindrical radio telescopes, UTMOST and CHIME. 
The formalism of the event rate prediction is followed by the predicted pulse widths 
of FRBs, the minimum total energy required to detect events and 
the redshift distribution of FRBs. We consider three different kinds of beam formations
for OWFA; viz. incoherent, coherent single and coherent multiple beam formations.

The predicted pulse widths of FRBs are estimated by considering two different 
scattering models and an arbitrary intrinsic pulse width. The contributions from 
residual dispersion smearing and cosmic expansion are also taken into account. 
We find that the intrinsic pulse width of FRB ($w_i$) makes 
insignificant contribution to the predicted pulse width for all redshifts, 
whereas scatter broadening however tend to dominate the pulse width. 
The pulse width vs $z$ variation is steeper for the scattering model I than 
that of scattering model II. The predicted pulse width for UTMOST 
and CHIME are smaller than that for OWFA, however the qualitative nature of 
the curves are similar.

The detection prospects of FRBs depend on the amount of the  
total energy received by the telescope from the source. We find the OWFA 
will be capable of detecting FRBs with $\alpha\gtrsim 0$. However, there is 
no cut off in redshift for UTMOST and CHIME, which is due to their higher operating 
frequencies.

We estimated the redshift distribution of the predicted FRBs for OWFA, 
UTMOST and CHIME. The qualitative nature of the curves are similar 
for all the three telescopes. We however find that the curves peak at a certain 
redshift ($z_p$) in the range $0.3\leq z_p \leq 1$ for scattering model I, 
whereas the curves are almost flat extending up to a high redshift for 
scattering model II. As a result the expected events per 
redshift range is low.

Finally, we have estimated the detection rates of FRBs for different phases of 
OWFA, considering different kinds of beam formation scenarios and compared them 
with the rates estimated for UTMOST and CHIME. The detection rates primarily 
depend on two important factors, the sensitivity and the field of view. 
The detection rate is higher for a telescope with a large field of view 
and a high sensitivity, and is qualitatively proportional 
to the product of these two factors. We find that the value 
of this product is maximum for OWFA Phase II in the coherent multiple 
beam formation mode. Our analysis predict that for OWFA Phase II, we can 
expect $\sim 100$ FRBs per day. The detection rates with the scattering model 
I are large compared to the that of scattering model II.

There are however some limitations of our analysis. The scattering mechanism 
in the intervening medium is still unknown. Cordes et al. (2016) showed that 
the scatter broadening is not proportional to DM for currently known sample of FRBs. 
Moreover, there is no unique and direct way to estimate the spectral index of 
FRBs. We have addressed this to a certain extent by considering both negative 
and as well as positive spectral indices. 
The energy distribution of FRBs is another important unknown and we have considered 
two possible energy distribution functions. Detection of FRBs in large numbers will 
therefore help us to constrain many of these uncertainties and refine the of FRBs 
models including their energy and redshift distributions.

\end{document}